\documentclass[useAMS,usenatbib]{mn2e}

\usepackage{graphics}
\usepackage{graphicx}
\title[Combined analysis of weak lensing and X-ray blind surveys]
{Combined analysis of weak lensing and X-ray blind surveys\thanks{Based on observations obtained with MegaPrime/MegaCam, a joint project of CFHT and CEA/DAPNIA, at the Canada-France-Hawaii Telescope (CFHT) which is operated by the National Research Council (NRC) of Canada, the Institut National des Sciences de l'Univers of the Centre National de la Recherche Scientifique (CNRS) of France, and the University of Hawaii. This work is based in part on data products produced at TERAPIX and the Canadian Astronomy Data Centre as part of the Canada-France-Hawaii Telescope Legacy Survey, a collaborative project of NRC and CNRS. It makes use of photometric redshifts produced jointly by TERAPIX and VVDS teams.}}
\author[J. Berg\'e et al.]{Joel Berg\'e $^1$\thanks{E-mail address : joel.berge@jpl.nasa.gov},  Florian Pacaud $^{1,2}$, Alexandre R\'efr\'egier $^1$, Richard Massey $^3$, \newauthor Marguerite Pierre $^1$, Adam Amara $^1$, Mark Birkinshaw $^4$, \newauthor St\'ephane Paulin-Henriksson $^1$, 
Graham P. Smith $^{5,3}$, Jon Willis $^6$ \\
$^1$ Laboratoire AIM, CEA/DSM - CNRS - Universit\'e Paris Diderot, DAPNIA/SAp, 91191 Gif-sur-Yvette, France \\
$^2$ Argelander Institute f\"ur Astronomy, Universit\"at Bonn, Auf dem H\"ugel 71, 53121 Bonn, Germany \\
$^3$ California Institute of Technology, 1200 E. California Blvd, Pasadena CA 91125, USA \\
$^4$ Department of Physics, University of Bristol, Tyndall Avenue, Bristol BS8 ITL, UK \\
$^5$ School of Physics and Astronomy, University of Birmingham, Edgbaston, Birmingham, B15 2TT, UK \\
$^6$ Department of Physics and Astronomy, University of Victoria, Elliot Building, 380 Finnerty Road, Victoria, V8V 1A1, BC, Canada}
\begin{document}

\date{Accepted . Received ; in original form }

\pagerange{\pageref{firstpage}--\pageref{lastpage}} \pubyear{}

\maketitle

\label{firstpage}
\setlength\tabcolsep{3pt}

\begin{abstract}
We present a joint weak lensing and X-ray analysis of 4 deg$^2$ from the CFHTLS and XMM-LSS surveys. Our weak lensing analysis is the first analysis of a real survey using shapelets, a new generation weak lensing analysis method. We create projected mass maps of the images, and extract 6 weak-lensing-detected clusters of galaxies. We show that their counts can be used to constrain the power spectrum normalisation $\sigma_8 =0.92_{-0.30}^{+0.26}$ for $\Omega_m=0.24$. We show that despite the large scatter generally observed in the M-T relation derived from lensing masses, tight constraints on both its slope and normalisation $M_*$ can be obtained with a moderate number of sources provided that the covered mass range is large enough. Adding clusters from \cite{bardeau07} to our sample, we measure $M_* = 2.71_{-0.61}^{+0.79} \, 10^{14} h^{-1} M_\odot$. Although they are dominated by shot noise and sample variance, our measurements are consistent with currently favoured values, and set the stage for future surveys. We thus investigate the dependence of those estimates on survey size, depth, and integration time, for joint weak lensing and X-ray surveys.  We show that deep surveys should be dedicated to the study of the physics of clusters and groups of galaxies. For a given exposure time, wide surveys provide a larger number of detected clusters and are therefore preferred for the measurement of cosmological parameters such as $\sigma_8$ and $M_*$. We show that a wide survey of a few hundred square degrees is needed to improve upon current measurements of these parameters. More ambitious surveys covering 7000 deg$^2$ will provide the 1\% accuracy in the estimation of the power spectrum and the M-T relation normalisations.
\end{abstract}


\section{Introduction}

In the currently-favoured hierarchical model of structure formation, clusters of galaxies have formed from the collapse of gravitational potential wells (e.g. \citealt{peebles80,padmanabhan93,lacey93,lokas01}) and are powerful cosmological probes. For instance, since they are sensitive to the expansion history of the Universe, their abundance and spatial distribution (e.g. \citealt{viana96,wang98,horellou05,nunes06,manera06}) and their mass function (e.g. \citealt{lokas04}) depend on cosmological parameters, such as the dark energy equation of state parameter $w$ (e.g. \citealt{basilakos03,maor05,basilakos07}), or the power spectrum normalisation $\sigma_8$ (e.g. \citealt{seljak02,pierpaoli03}).
Several observational methods now permit the use of clusters of galaxies as cosmological probes, such as X-ray observations, weak gravitational lensing and the Sunyaev-Zeldovich effect. 

Due to improvements in telescopes and techniques, X-ray studies are able to constrain cluster physics and the mass scaling relation with ever greater precision. For instance, the self-similarity for clusters of galaxies (\citealt{eke98,arnaud02}) has been observationally verified. 
Nevertheless, adiabatic simulations still predicts a mass-temperature relation with double the observed normalisation (e.g. \citealt{nevalainen00,finoguenov01}) and the self-similarity assumption could break down at low temperatures ($T \leqslant 3 ~{\rm keV}$). Thus, a steepening of the mass-temperature (M-T) relation is expected if galaxy groups underwent a preheating by supernovae, or a surge of entropy, in their early days (\citealt{bialek01,muanwong02}). Recent evidence for this steepening was found by e.g. \cite{nevalainen00,finoguenov01}, or Arnaud, Pointecouteau \& Pratt 2005 (APP05 hereafter), but could not be seen by e.g. \cite{etorri02,castillo03,vikhlinin06}. 
Moreover, the M-T normalisation estimation is currently limited by systematics in measuring cluster masses from their X-ray profiles.
This limitation can be lifted by using probes which are independent of the physical state of the cluster. 

Beyond galaxy cluster physics, the M-T relation is needed by X-ray experiments to estimate the power spectrum normalisation $\sigma_8$. 
Measuring this parameter has triggered much effort in several observational areas. For instance, CMB experiments (e.g. \citealt{wmap3}) tend to a low value for $\sigma_8$ ($\leqslant 0.8$), weak lensing experiments tend to higher values ($\geqslant 0.8$), and X-ray observations provide intermediate values.

Gravitational lensing  does not depend on the underlying physics of clusters of galaxies or dark matter, but only on their potential wells, and on the Universe's geometry. 
Strong gravitational lensing has been used for galaxy clusters physics (e.g. \citealt{mellier93,kneib95,kneib96,smail97,luppino99,smith05}) and measurement of $\sigma_8$ (e.g. \citealt{smith03}).
Weak gravitational lensing is more difficult to measure (for reviews, see e.g. \citealt{mellier99,bartelmannschneider,alexreview,munshi06})), and has taken longer to be detected (\citealt{bacon00,vanwaerbeke00,wittman00,rhodes01}). Since then, particular attention has been given to cosmic shear, i.e. statistical cosmological weak lensing (e.g.  \citealt{bacon02,heymans04,massey04,cosmos,hoekstra06,semboloni06,schrabback06}), in attempts to measure $w$ and $\sigma_8$. It has also begun to be used as a tracer of the cosmic web (e.g. \citealt{cosmosmaps}), and a way to detect and catalogue mass overdensities (e.g. \citealt{wittman06}, Gavazzi \& Soucail 2007, GS07 hereafter, and \citealt{miyazaki07}). 
Beside the constraints it can bring to cosmology, it can be used as a complement to X-ray analyses of clusters of galaxies.
Thanks to the physics-independent estimation of cluster masses, it appears as a unique method to calibrate the mass-temperature relation for clusters of galaxies (e.g. \citealt{hjorth98,huterer02,pedersen06,bardeau07}). It has been shown that the uncertainty in the normalisation of the mass-temperature relation is the largest source of error in $\sigma_8$ measurements inferred from X-ray cluster analyses (\citealt{seljak02,pierpaoli03}). An accurate mass-temperature relation, obtained from combined weak lensing and X-ray analyses, will thus provide new insights not only on the $\sigma_8$ discrepancy, but also on galaxy cluster physics. 

In this paper, we present the first joint analysis of weak gravitational lensing and X-ray wide-area surveys of a randomly-selected patch of sky. The weak lensing survey is derived from the CFHTLS, and the X-ray survey from the XMM-LSS. The weak lensing analysis uses shapelets (\citealt{shapelets1,shapelets2,shapelets3}), a new generation shear measurement technique, which has been shown to achieve a few percent accuracy in shear measurement from ground based telescopes (\citealt{step2}). We analyse one square-degree of the CFHTLS Deep survey (the D1 field) and four contiguous square degrees of the CFHTLS Wide survey, which enclose the D1 field. We create convergence maps for this region of the sky and give a catalogue of detections.
We show how counting weak-lensing-selected clusters can provide an estimate of the power spectrum normalisation $\sigma_8$.
We then show how the combination of weak lensing and X-ray analyses of clusters provides an estimate of the mass-temperature relation normalisation $T_*$, independent of clusters physical state.
Finally, we investigate the impact of a joint weak lensing and X-ray survey strategy on the accuracy of the $\sigma_8$ and $T_*$ measurement. We consider deep and wide weak lensing surveys, with the CFHTLS characteristics, combined with a blind X-ray survey of the same region of the sky.

The organisation of the paper is as follows. Section \ref{data} presents the surveys used in this paper, namely the CFHTLS and the XMM-LSS. The methods that we use are described in section \ref{method}. We show how we estimate the weak lensing effect using shapelets, and how we generate convergence maps and detect clusters. We also briefly describe how the X-ray properties of clusters are obtained. Section \ref{results} presents the convergence map we inferred from our weak lensing analysis, and gives a catalogue of the galaxy clusters that we detect. We then give our estimates of the normalisation of the power spectrum and of the M-T relation. We then show in section \ref{discussion} that combined blind surveys are necessary to get the best insights about those normalisations. The impact of survey strategy on those parameters estimations is discussed in section \ref{impact}. We conclude in section \ref{conclusion}.

\section{Data} \label{data}

\subsection{Weak lensing : CFHTLS}

The ``Canada-France-Hawaii Telescope Legacy Survey" \footnote{http://www.cfht.hawaii.edu/Science/CFHLS/} (CFHTLS), a joint France-Canada project, consists of three different surveys, namely the Very Wide Survey, the Wide Synoptic Survey (referred to as ``Wide survey" hereafter), and the Deep Survey. 
Once complete, the Wide Survey will cover 170 deg$^2$ (divided into four distinct patches ranging from 49 deg$^2$ to 72 deg$^2$) in five filters (u*,g',r',i',z'), down to a magnitude ${\rm i}' \approx 24.5$. Its main goal is the study of large scale structures by weak gravitational lensing and galaxy counts. The Deep Survey covers 4 different fields, each with an area of 1 deg$^2$, in the same five filters, down to ${\rm i}' \approx 28.5$. It is primarily intended for Type Ia Supernovae studies but it is also useful for measuring  large-scale structures. The CFHTLS images were obtained from observations with the MegaCam camera, made of a 36 CCD mosaic, of 2048 $\times$ 4196 pixels each, with a 1 deg$^2$ field of view (\citealt{boulade03}).

\begin{figure}
\centering
\includegraphics[width=8cm,angle=0]{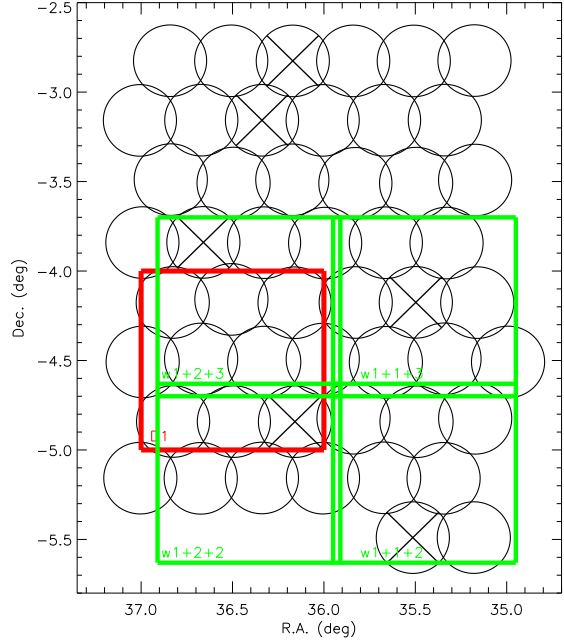} 
\caption{Layout of the surveys. The red square is the CFHTLS D1 field. The four green squares are the four CFHTLS W1 fields used in this paper. Circles represent the XMM-LSS pointings available in that region, prior to XMM AO5 (\citealt{pacaud07}). Here, we only used those lying within the optical data. Those marked by a cross are strongly affected by flares and are being re-observed.} \label{fig_data}
\end{figure}

In this paper, we present the weak gravitational lensing based mapping of 4 deg$^2$ of the Wide Survey (W1 patch), which include the 1 deg$^2$ field of the Deep Survey (D1 field), using both W1 and D1 images. 
The geometry of the fields that we use is shown in Figure \ref{fig_data}.
The data processing (astrometry, photometric calibration, stacking of images) has been done by the CFHT community and Terapix \footnote{http://terapix.iap.fr}. 
We use W1 images optimised for weak lensing : each field is the combination of 7 stacked images, each of 620 seconds exposure time. We use the T0003 release of the D1 field, consisting of 275 stacked images, with a total 37.4 hour exposure time. The average seeing is 0.7 arcsec.
We masked parts of the images with saturated stars and/or too high a noise, by hand, so as not to bias our weak lensing results. This operation removes 10\% of the original area covered by the data. We do not mask the ghosts created by spurious reflections on the telescope optics around saturated stars, but we eventually remove the galaxies that they cover from our catalogues, since they are too noisy. 
The average galaxy density is 28 arcmin$^{-2}$ in the D1 image, and 13 arcmin$^{-2}$ in the W1 images.

\subsection{X-ray : XMM-LSS}

Designed to cover an area of several tens of square degrees up to a redshift $z=1$, the XMM-LSS survey aims at detecting a significant fraction of clusters of galaxies, in order to constitute a sample suited to cosmological studies (\citealt{pierre04}). Its nominal exposure times are 10 ks, and have been raised up to 20 ks for the {\it XMM} Medium Deep Survey (XMDS : \citealt{chiappetti05}), a 2 deg$^2$ region included in the XMM-LSS, which covers the CFHTLS D1 field. In this paper, we use 4 deg$^2$ of the XMM-LSS which cover our 4 deg$^2$ CFHTLS data. The {\it XMM} pointings are shown on Figure \ref{fig_data}. The raw X-ray observations reduction is presented in \cite{pacaud06}.

\section{Method} \label{method}

\subsection{Weak lensing cluster detection}

Introduced in \citealt{shapelets1}, \citealt{shapelets2} and \citealt{shapelets3}, shapelets are a complete, orthogonal, set of basis functions with which one can analytically decompose galaxy shapes. They can be seen as fundamental shapes : a particular galaxy can be represented by a particular sum of shapelets basis functions $\chi_{n,m}$, weighted by coefficients $f_{n,m}$. Their rich formalism provides an intuitive and analytic form for geometrical transformations (such as smear, shear, rotation) and for (de)convolution. Hence, they allow one to analytically describe the smearing of the Point Spread Function (PSF) and the shear of galaxies, properly correcting for the PSF. The shear estimation they provide has been shown to reach the needed accuracy for the CFHTLS specifications by the STEP project (\citealt{step2}). 

Our full pipeline will be described in an upcoming paper (Berg\'e et al. in prep). We briefly summarise it here.
Each sufficiently bright and non-saturated star is first decomposed into shapelets. A polynomial interpolation of each shapelet coefficient then provides a model of its spatial variations across the image. We are thus able to reconstruct the shape of the PSF at the position of each galaxy, the condition necessary for deconvolving it from the galaxies' shapes. Several stringent tests then validate our PSF model. In particular, we require that the ellipticity distribution, and the two point correlation functions of the ellipticity, of the residuals between observed stars and their shapelet models are consistent with zero. We also require that the cross-correlation between stars and galaxies ellipticity, when corrected from the PSF, is consistent with zero. The shape of galaxies is finally measured by decomposing them into shapelets, while deconvolving from the PSF, as shown in \citealt{shapelets3}.

A shear estimator is created from the shapelet decompositions of galaxies, as prescribed by \citealt{shapelets_shear}:

\begin{equation} \label{gamma}
\gamma = \frac{f_{2,2}}{P^\gamma}
\end{equation}
where the shear susceptibility $P^\gamma = <f_{0,0}-f_{4,0}> / \sqrt{2}$ is fitted on the magnitude-size plane for galaxies. The coefficients are complex numbers. The shear $\gamma$ of equation (\ref{gamma}) is the complex notation for shears, $\gamma = \gamma_1+i \gamma_2$.

\begin{figure*}
\centering
\includegraphics[width=11cm,angle=0]{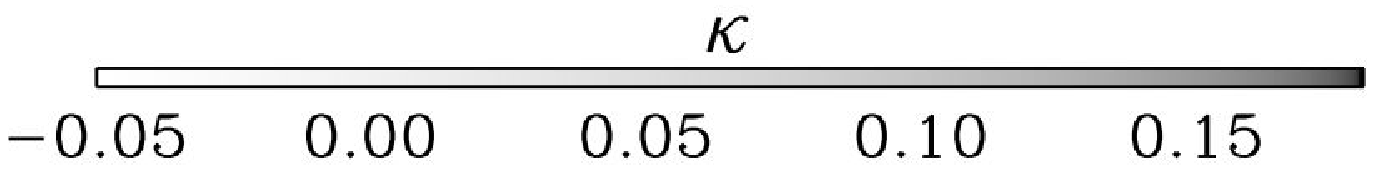} 
\includegraphics[width=12cm,angle=0]{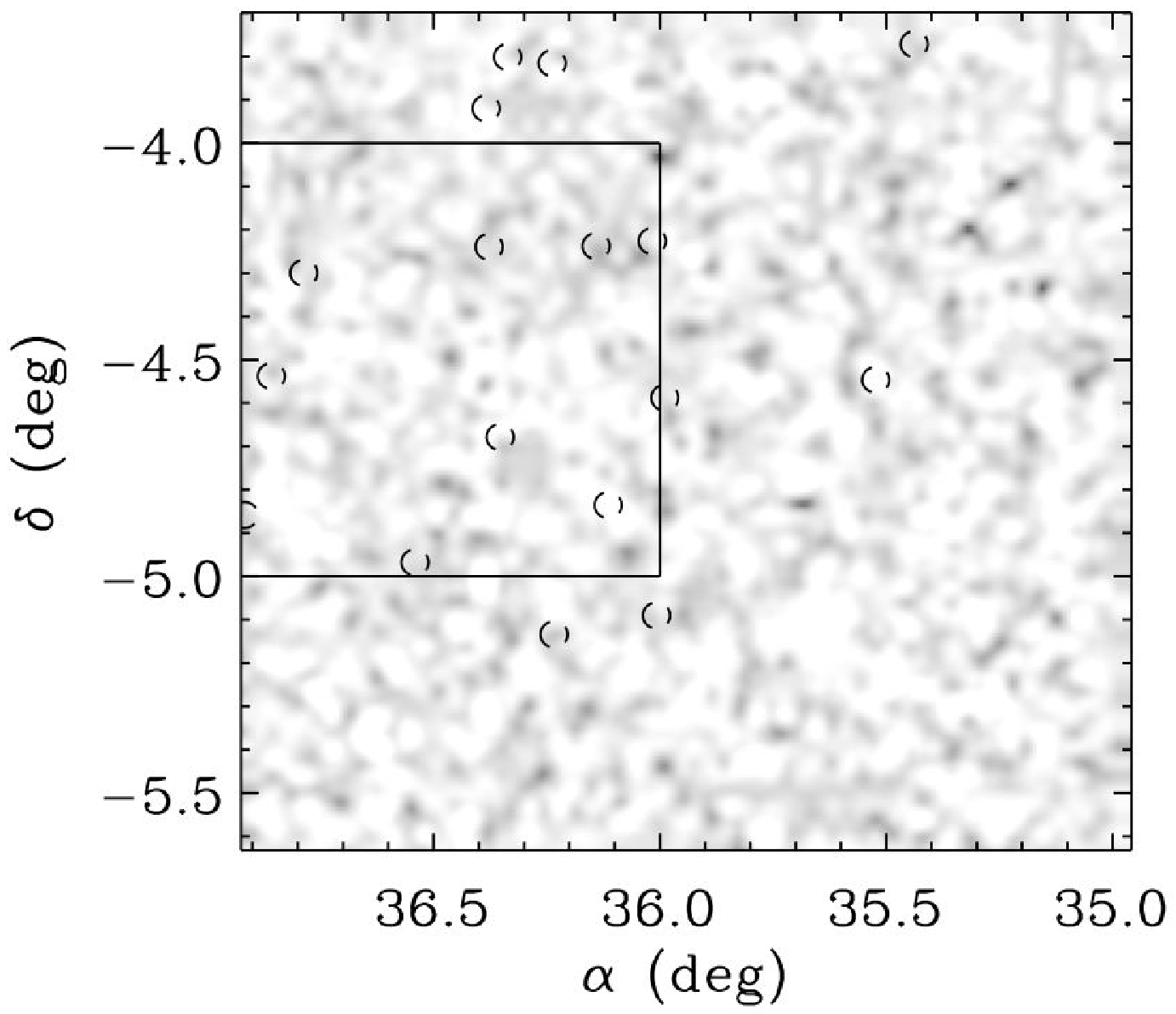}
\caption{Convergence map inferred from our weak lensing measurement of the W1 field. The square in the W1 map shows the boundaries of the D1 field (Fig. \ref{massmap}). The map is smoothed by a 2.3' FWHM Gaussian. Dashed circles mark C1 X-ray clusters.  } \label{massmapw1}
\end{figure*}

To increase the signal to noise ratio of our measurements, we give to each galaxy $g$ a minimum variance weight $w_g = (\sigma_{\varepsilon,g}^2 + \sigma_{P_\gamma,g}^2 + \sigma_{\rm int}^2)^{-1}$, where $\sigma_{\varepsilon,g}$ is the error on shape measurement for galaxy $g$, $\sigma_{P_{\gamma,g}}$ the error on the measurement of its shear susceptibility, and $\sigma_{\rm int}$ is the intrinsic ellipticity dispersion, set to $\sigma_{\rm int}=0.3$. 
Slightly changing $\sigma_{\rm int}$ would be equivalent to giving more or less weight to our measurement errors, and would mostly affect the error bars in the shear measurement. The eventual peak detection would not be affected by such slight changes.
This weighting scheme is equivalent to smoothly selecting the most useful galaxies for shear measurement. For instance, the faintest are down-weighted. 
It therefore provides us with effective densities of $n_{\rm eff} \approx 20$  and $9$ useful galaxies per square arcminute, in the D1 and W1 images, respectively.
Then, a direct inversion in Fourier space of the pixelised shear map  allows us to infer a convergence (i.e. projected mass) map (\citealt{ks93}) of the images.
Structures in this mass map are extracted from the noise using  a Gaussian filter. Figures \ref{massmapw1} and \ref{massmap} show the convergence maps that we inferred from our data. These are described in section \ref{catalogue}. While constructing a convergence map, we also create a signal-to-noise map, the signal-to-noise ratio being defined as 

\begin{equation}
\nu (x,y) = \frac{\kappa (x,y)}{\sigma_\kappa (x,y)}, 
\end{equation}
where $\kappa (x,y)$ is the convergence at the ($x,y$) sky coordinates, and $\sigma_\kappa (x,y)$ its r.m.s error. The r.m.s error $\sigma_\kappa (x,y)$ is computed using Monte-Carlo simulations in which the input galaxies are positioned like the observed ones but with randomised shape orientations. 

Structures are then searched for in the filtered convergence map, and their astrometry provided, by the SExtractor software (\citealt{sextractor}). They are extracted according to their signal-to-noise peak, read from the signal-to-noise map. Hereafter, we define a `significant structure' as a detection with a signal-to-noise ratio greater than 2.5.
Their mass is related to their integrated convergence through the lensing geometry, and can be estimated when their redshift and the redshift distribution of background galaxies are known. To account for the latter, we use  the normalised distribution

\begin{equation} \label{eq_nz} 
n(z)=\frac{\beta}{z_s \Gamma (\frac{1+\alpha}{\beta})} \left( \frac{z}{z_s} \right) ^\alpha \exp \left[ -\left(\frac{z}{z_s} \right) ^\beta \right]
\end{equation}
where the parameters $\alpha,\beta,z_s$ are given for the Wide images by \cite{benjamin07} $(\alpha,\beta,z_s)=(0.836,3.425,1.171)$.
To account for $n(z)$ in the D1 image, we fit \cite{ilbert06}'s photometric redshift distribution in the CFHTLS D1 field, and obtain $(\alpha,\beta,z_s)=(0.828,1.859,1.148)$. \cite{vanwaerbeke06} have shown that errors in the $n(z)$ fit are subdominant compared to Poisson noise and sample variance for the measurement of cosmological parameters. We thus neglect them hereafter.

We measure a cluster's virial mass by averaging its convergence in an aperture large enough that we can assume that the entire cluster is captured. The aperture corresponds to the region enclosed in the 2$\sigma$ level of the cluster's convergence map. This technique is similar to using a $\zeta$-statistic (\citealt{fahlman94}), with infinitely large annulus around the cluster, provided that the convergence in the entire field averages to 0. We verified this latter point, thus validating our choice. Note that because of the small number density of background sources, we cannot reliably fit a shear profile around clusters (see e.g. \cite{paulin07} for an example of mass estimation using two profile fits around the galaxy cluster Abell 209). We then  convert the virial mass into $M_{200,c}$, the mass enclosed in the sphere of mean overdensity 200 times higher than the critical density, using the recipe from \cite{hu_kravtsov}. Hereafter, we will note $M_{200,c}$ more briefly $M_{200}$.

Weak lensing is affected by the entire mass distribution along the line of sight. As a result, the weak lensing mass measurement of one cluster is biased by projection effects. It has been shown, using different mass estimators, that large-scale structures in the line of sight, and near the target cluster, introduce errors ranging from a few percent (\citealt{reblinsky99,hoekstra01,hoekstra03,clowe04}) to a few tens of percent (\citealt{metzler01,deputter05}). 
In this paper, we assume that they produce a 20\% error, added in quadrature to the shear measurement error.

\subsection{X-ray cluster detection and analysis}

The X-ray cluster detection pipeline has been described in \cite{pacaud06}. It takes account of the Poisson nature of the X-ray images, to extract and analyse clusters of galaxies in a two-step procedure. Clusters are first detected by a multi-resolution wavelet filter (\citealt{starck98}). Then, each source is analysed using a maximum likelihood profile fitting procedure, and its X-ray properties assessed. Three classes of extended sources have been defined (\citealt{pacaud06,pierre06}) : (1) the C1 class contains the highest surface brightness sources, and is uncontaminated ; (2) the C2 class allows for 50 \% contamination, and contains less bright extended sources ; (3) finally, the C3 class contains optically confirmed sources with apparent X-ray emission, that were not selected as C1 or C2. In this paper, we only consider C1 class detections, representative of the most massive objects seen in the XMM-LSS.
The redshift of detected clusters has been measured using spectroscopic observations from a number of telescope and instrument combinations detailed in Table 2 of \cite{pierre06}. Their temperature estimation is described in \cite{willis05}.

\cite{pacaud07} have extracted and analysed 29 C1 clusters from 5 deg$^2$ of the XMM-LSS data (shown on Fig. \ref{fig_data}), that contain our 4 deg$^2$ optical data. Among other things, they have measured their luminosity and temperature. Here, we take into account their 16 clusters which are enclosed in the fields of our CFHTLS data, making use only of their temperature and redshift. They are listed in Table \ref{cataloguetable}. Note that the cluster XLSSC053 is in the G12 XMM-LSS pointing (shown by the cross in the D1 field, on Fig. \ref{fig_data}), which was not used when \cite{pacaud07} analysed the XMM-LSS observations. This pointing has been re-observed, and the X-ray characteristics of XLSSC053 are listed in Table \ref{cataloguetable}.

\section{Results} \label{results}

In this section, we give the properties of clusters of galaxies detected with our weak lensing pipeline. Counting the weak lensing detections allows us to constrain the matter power spectrum normalisation $\sigma_8$. 
We then use the weak lensing mass of the detected groups to calibrate the mass-temperature relation for clusters of galaxies.

\subsection{Convergence maps and cluster catalogue} \label{catalogue}


\begin{figure*}
\centering
\includegraphics[width=11cm,angle=0]{fig3a.eps} 
\includegraphics[width=12cm,angle=0]{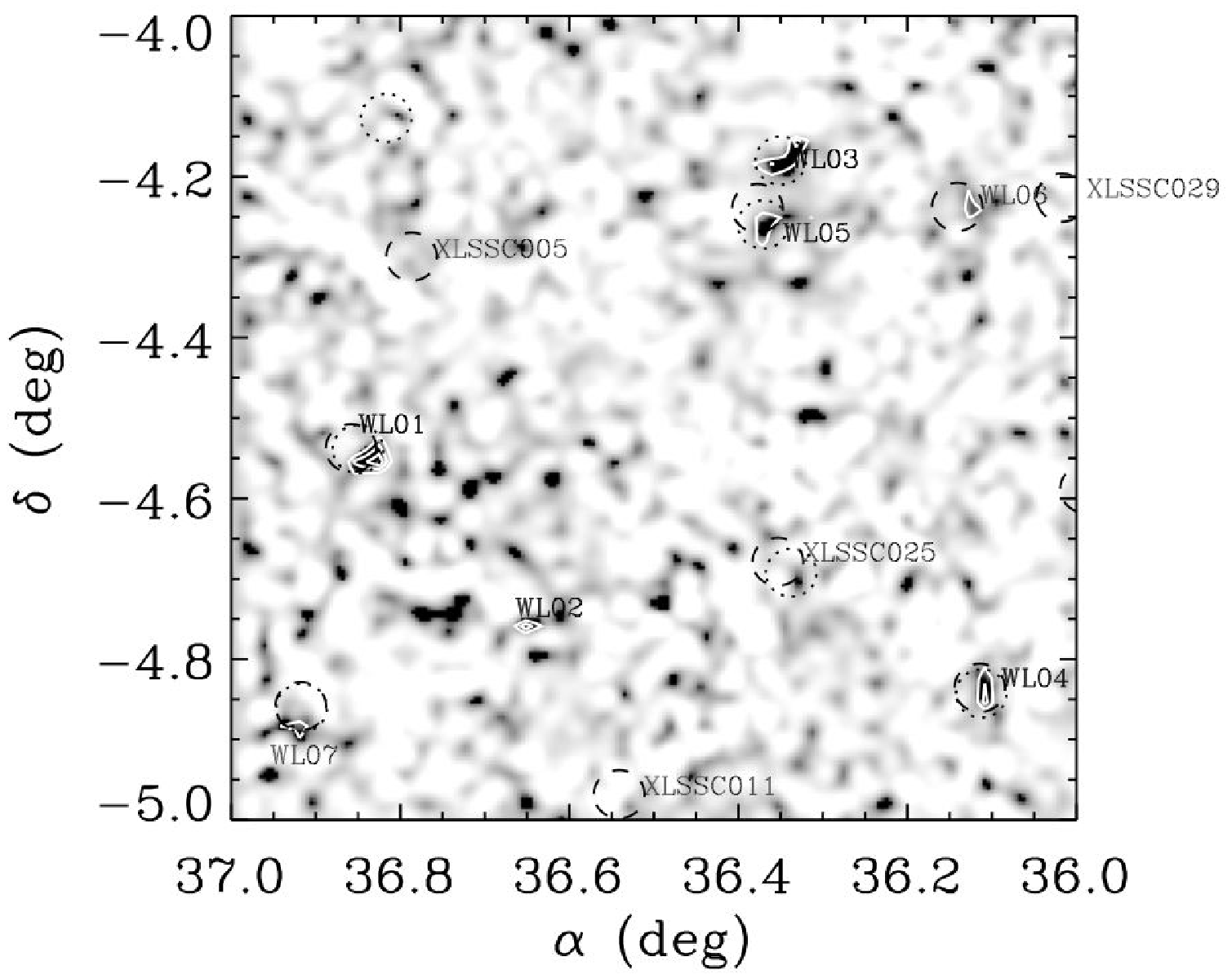}
\includegraphics[width=6cm,angle=0]{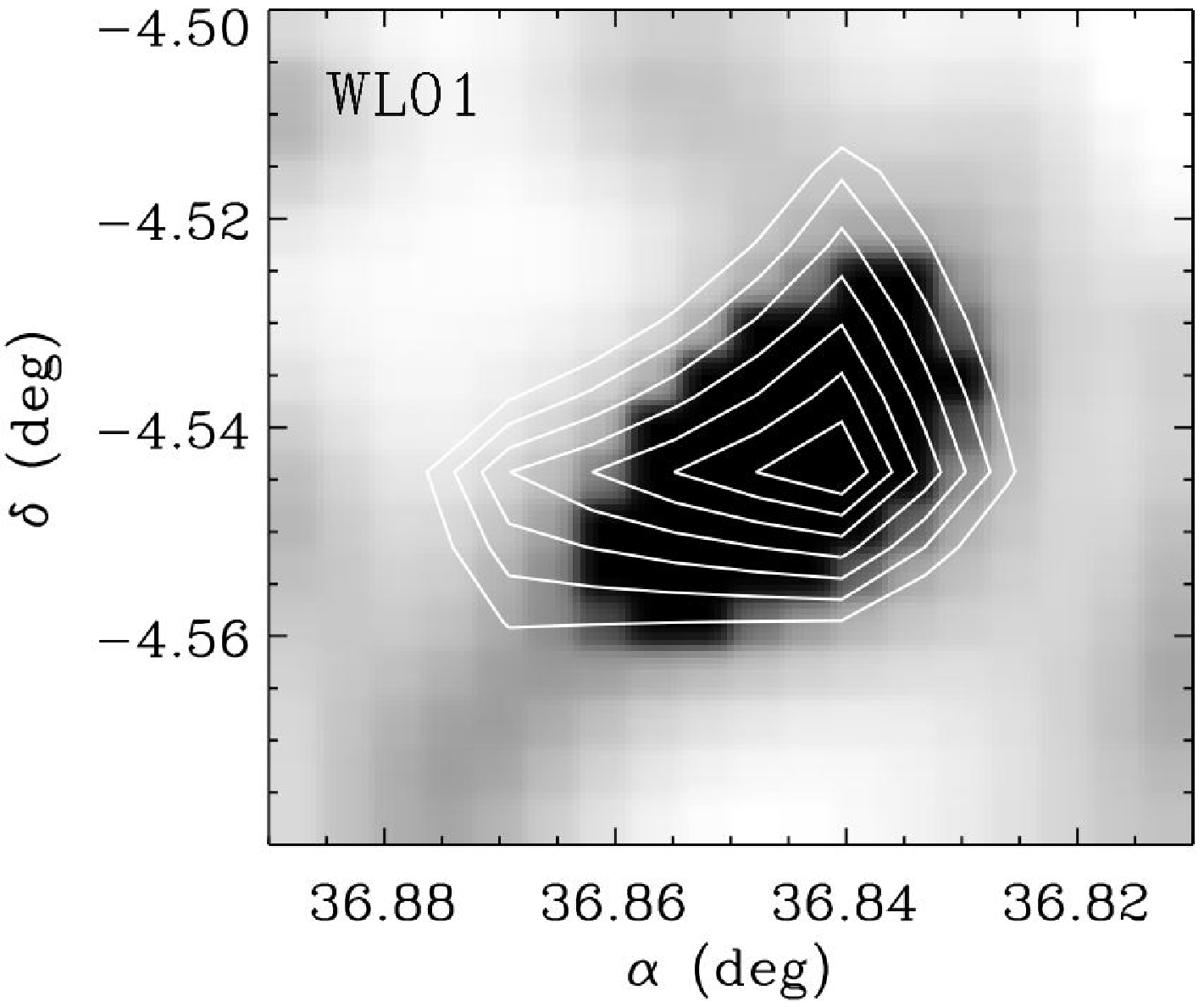}
 \includegraphics[width=6cm,angle=0]{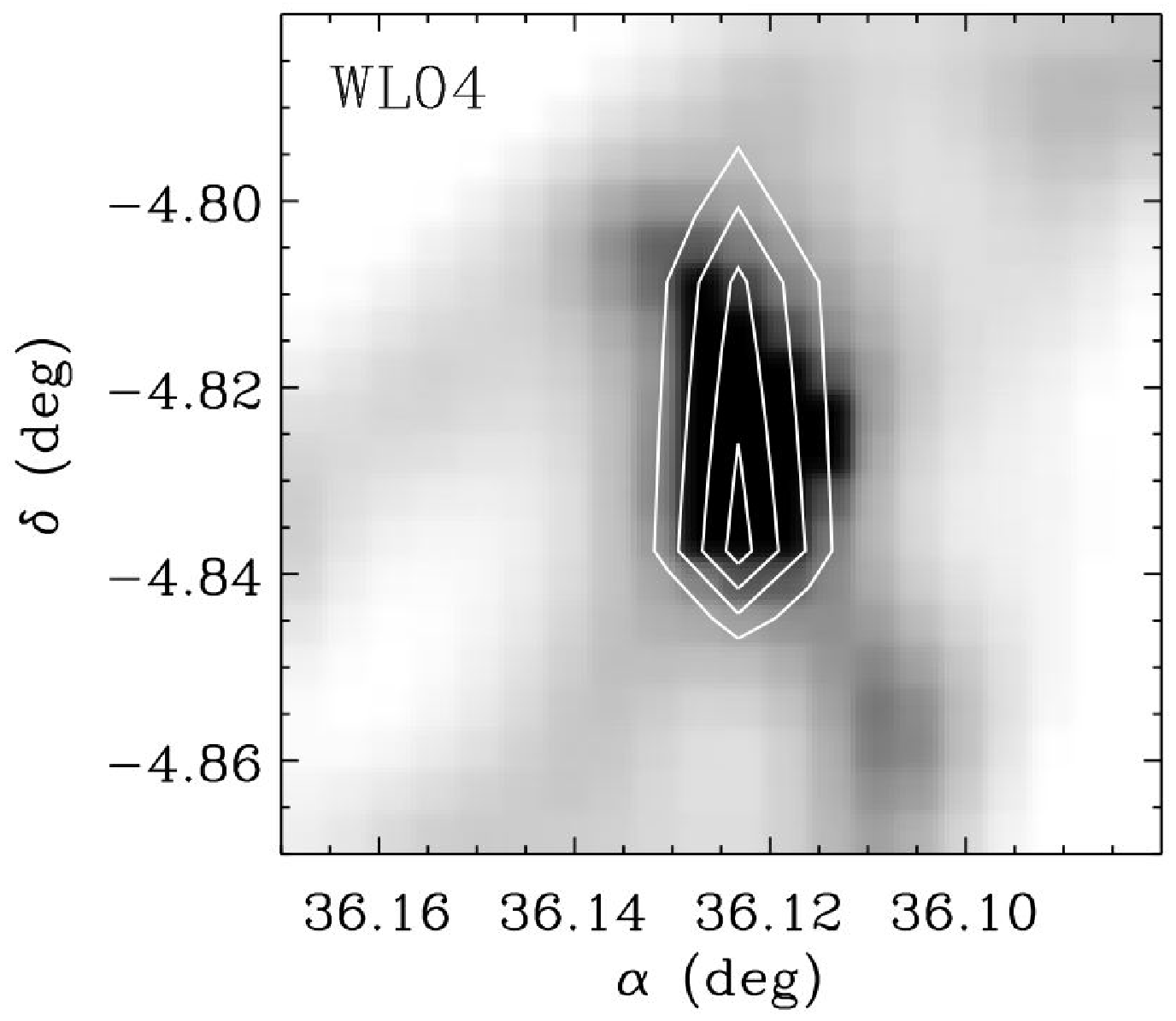} 
 \includegraphics[width=11cm,angle=0]{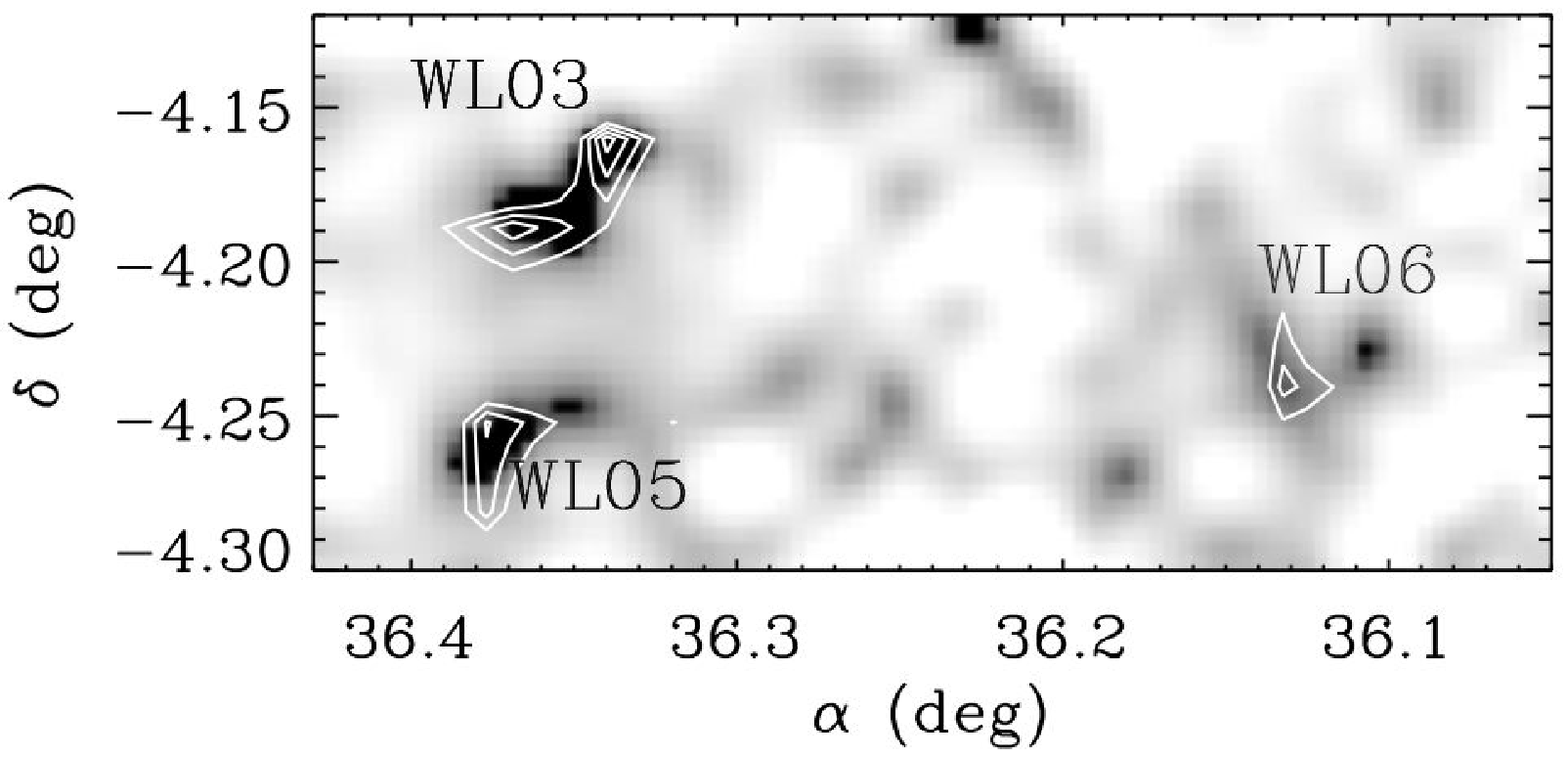} 
\caption{Convergence map inferred from our weak lensing measurement of the D1 field. The convergence $\kappa$ is shown for the entire field on the upper panel. Lower panels show zooms on cluster candidates WL01, WL04, and the region surrounding WL03, WL05 and WL06. In the upper panel, contours levels start at 2.2$\sigma$, with an increment of 0.5$\sigma$. In the lower panels, they start at 2.2$\sigma$, with an increment of 0.2$\sigma$. The maps are smoothed by a 1.1' FWHM Gaussian. In the upper panel, dashed circles mark X-ray clusters, and dotted circles show \citealt{gavazzi}'s KSB weak lensing detections. Clusters detected by our shapelets weak lensing measurement are labelled WL{\it id}, and X-ray clusters not detected by weak lensing are labelled by their {\it XMM} name. All clusters are listed in Table \ref{cataloguetable}. For clarity, false detections near edges are not shown.} \label{massmap}
\end{figure*}

\begin{table*}
\caption{Clusters catalogue. Besides the clusters that we detect through our shapelet weak lensing measurement, we also list clusters seen by \citealt{gavazzi}, and selected as C1 clusters in the XMM-LSS. 
Weak lensing detections' significance is given for D1 and W1, even if no detection appears in W1.
$M_{200}$(WL) is the cluster weak lensing mass. $M_{200}$(X) is the mass estimated from X-ray profile, extrapolated from $R_{500}$ to $R_{200}$, with respect to $M_{500}$ masses given by Pacaud et al. (2007), and must be used with caution (see text).}
\begin{center}
\begin{tabular}{|c|c|c|c|c|c|c|c|c|c|c|}
\hline
Weak lensing & XLSSC & GS07 & RA & Dec & $z$ & Significance & $M_{200}$(WL) & $T_X$  & $M_{200}$(X)$^c$ & Notes \\
 cluster ID & number & ID & (deg) & (deg) & & D1/W1 & ($10^{13} h^{-1} M_\odot$) & (keV)  &  ($10^{13} h^{-1} M_\odot$) \\
\hline
WL01 & 013 & Cl03 & 36.8497 & -4.5481 & 0.31 & 3.61 / - & $8.2_{-1.9}^{+2.5}$ & $1.0_{-0.1}^{+0.1}$  & 2.1 \\
WL02 & - & - & 36.6589 & -4.7516 & - & 3.09 / - & - & - & -  \\
WL03 & - & Cl04 & 36.3628 & -4.1886 & 0.32 $^a$ & 2.91 / - & $8.9_{-2.2}^{+2.6}$ & - & -  \\
WL04 & 053 & Cl02 & 36.1229 & -4.8341 & 0.50 $^b$ & 2.90 / - & $10.3_{-2.6}^{+3.0}$ & $3.4_{-1.0}^{+3.1}$  & 5.0 & XMM-LSS pointing \\
 & & & & & & & & & & not observed in \\
& & & & & & & & & & \cite{pacaud07} \\ 
WL05 & 041 & Cl14 & 36.3723 & -4.2604 & 0.14 & 2.62 / - & $4.9_{-1.2}^{+1.6}$ & $1.3_{-0.1}^{+0.1}$ & 3.5 \\
WL06 & 044 & - & 36.1389 & -4.2384 & 0.26 & 2.48 / - & $7.2_{-1.7}^{+2.3}$ & $1.3_{-0.1}^{+0.2}$  & 3.7 & just below detection threshold \\
 & & & & & & & & & & in GS07's catalogue \\
WL07 $^d$ & 022 & Cl07 & 36.9167 & -4.8606 & 0.29 & 2.42 / - & - & $1.7_{-0.1}^{+0.1}$  & 5.3 & near a mask   \\
- & 025 & Cl05 & 36.3375 & -4.6925 & 0.26 & - / - & - & $2.0_{-0.2}^{+0.2}$  & 6.5 & under a mask   \\
- & - & Cl10 & 36.8167 & -4.1269 & - & - / - & - & - \\
- & 029 & - & 36.0172 & -4.2260 & 1.05 & - / - & - & $4.1_{-0.7}^{+0.9}$   & 13.9 & too high redshift \\
- & 011 & - & 36.5410 & -4.9680 & 0.05 & - / - & - & $0.64_{-0.04}^{+0.06}$   & 1.0 & \\
- & 005 & - & 36.7866 & -4.2995 & 1.05 & - / - & - & $3.7_{-1.}^{+1.5}$  & 16.5  & too high redshift \\
- & 006$^\dagger$ & - & 35.4382 & -3.7717 & 0.43 & X / - & - & $4.8_{-0.5}^{+0.6}$  & 30.4 & near an edge \\
- & 040$^\dagger$ & - & 35.5232 & -4.5463 & 0.32 & X / - & - & $1.6_{-0.3}^{+1.1}$  & 6.8 \\
- & 049$^\dagger$ & - & 35.9892 & -4.5880 & 0.49 & X / - & - & $2.2_{-0.5}^{+0.9}$  & 5.0 \\
- & 018$^\dagger$ & - & 36.0079 & -5.0903 & 0.32 & X / - & - & $2.0_{-0.4}^{+0.7}$  & 8.0 \\
- & 021$^\dagger$ & - & 36.2338 & -5.1340 & 0.08 & X / - & - & $0.68_{-0.02}^{+0.04}$  & 1.8 \\
- & 001$^\dagger$ & - & 36.2378 & -3.8156 & 0.61 & X / - & - & $3.2_{-0.3}^{+0.4}$  & 14.3 \\
- & 008$^\dagger$ & - & 36.3367 & -3.8014 & 0.30 & X / - & - & $1.3_{-0.2}^{+0.7}$  & 2.1 \\
- &  002$^\dagger$ & - & 36.3841 & -3.9198 & 0.77 & X / - & - & $2.8_{-0.5}^{+0.8}$  & 9.6 \\
\hline
\multicolumn{11}{l}{$^a$ Tomographic redshift (\citealt{gavazzi})} \\
\multicolumn{11}{l}{$^b$ Photometric redshift (Aussel et al. in prep)} \\
\multicolumn{11}{l}{$^c$ Rough estimates based on the isothermal assumption and extrapolated from $M_{500}$ given by \cite{pacaud07}.} \\
\multicolumn{11}{l}{$^d$ X-ray coordinates.} \\
\multicolumn{11}{l}{$^\dagger$ Outside D1.}  \\
\end{tabular}
\end{center}
\label{cataloguetable}
\end{table*}

Figure \ref{massmapw1} shows the 4 deg$^2$ of the W1 field that we considered. No significant overdensity (i.e. with $\nu \geq 2.5$) has been detected. 
As we will quantitatively show in section \ref{selfct}, this is consistent with the expected cluster counts for this survey. The black square in the image shows the position of the D1 field. Since there are around 20 galaxies per arcmin$^2$, we expect more significant detections in this deep field (see section \ref{selfct}).
The upper panel of Fig. \ref{massmap} shows the convergence map that we obtained from our weak lensing analysis of the D1 field.  Due to the varying level of noise in our map, which varies independently of $\kappa$, two peaks with the same $\kappa$ value do not necessarily have the same significance. That results in the rejection of seemingly significant structures, such as the peak around ($\alpha,\delta$)=($36.75^{\rm o},-4.75^{\rm o}$). 
Significant structures are marked out by the white contours, which start at 2.2$\sigma$, with an increment of 0.5$\sigma$. The lower panels show individual candidate clusters in more details. In these, contours start at 2.2$\sigma$, with an increment of 0.2$\sigma$. Even though we consider as significant a structure with $\nu \geqslant 2.5$, we plot the 2.2$\sigma$ contours as a way to show the extension of our detections.
Dashed circles mark X-ray clusters, and dotted circles \cite{gavazzi}'s weak lensing detections. \cite{gavazzi} measured the shear in the D1 field using the KSB method (\citealt{ksb}).

Table \ref{cataloguetable} summarises the measured characteristics of the clusters that we detect, together with all X-ray and \cite{gavazzi}'s detections in the region. 
The clusters that we detect through our shapelets weak lensing analysis are labelled with WL{\it id}, where {\it id} runs from 00 to 07, and are sorted by decreasing significance. Their labels are listed in the first column. 
Their official {\it XMM} names are given in column (2), and \cite{gavazzi}'s IDs in the third column.
X-ray clusters marked by a $^\dagger$ are outside the D1 field.
Columns (4) and (5) give their position.  Column (6) lists their spectroscopic redshifts, except for clusters WL03, for which a tomographic redshift is given, and WL04 for which a photometric redshift is given (Aussel et al. in prep). 
The significances of the weak lensing detections are listed in column (7), in D1 and W1. A `-' means that the cluster is not detected ; a `X' means that the cluster is outside the D1 field.
Columns (8) and (9) give their weak lensing mass $M_{200}$(WL) and X-ray temperature, respectively. 
Column (10) gives the mass estimate from the X-ray data, $M_{200}(X)$. As in \cite{pacaud07}, these were evaluated under the assumption of an 
isothermal $\beta$-model gas distribution in hydrostatic equilibrium with the cluster's potential well. In this earlier work, the associated statistical errors were generally dominated by the temperature measurement uncertainty (${\delta}M/M\approx{\delta}T/T$ of order 10-25\%). Here, the error on the emission profile can also become quite significant because we estimate the masses within $R_{200}$ instead of $R_{500}$ where the X-ray emission starts to vanish. Moreover, it was shown by \cite{vikhlinin06} and \cite{rasia06} that the isothermal $\beta$-model assumption leads to an underestimation of the total mass, by up to 40\% for low mass systems. For these reasons, we decided not to provide error bars for our X-ray masses.
Finally, column (11) gives some details about weak lensing detections, explaining for instance why we chose not to take them into account, or why we do not detect a cluster seen by another method. Among the rejection criteria are the proximity to an edge or to a masked region, the mass inversion procedure being sensitive to missing data and to edge effects. In that sense, and for clarity, the detections closest to edges have been removed from Fig. \ref{massmapw1} and Fig. \ref{massmap}. 

Clusters WL01, WL04 and WL05 have unequivocal counterparts both in our X-ray catalogue and in \citealt{gavazzi}'s KSB one. No significant C1 X-ray source has been selected around WL02, and it remains invisible to \citealt{gavazzi}. Moreover, a visual inspection of the optical images does not show any galaxy overdensity around it. No significant C1 X-ray source has been detected at the position of  WL03, even though it is also seen by \cite{gavazzi}. 
Cluster WL06 lies just below our detection threshold ($\nu =2.48$). Nevertheless, since \cite{gavazzi} detect it (though just below the detection threshold they use for their analyses) and since it is also detected by our X-ray analysis, we decided to list it, and to assess its weak lensing characteristics.
The significance of cluster WL07 is even lower ($\nu =2.42$). Since it is found close to the XLSSC022 cluster (which coincides with \cite{gavazzi}'s Cl07), we show its contours in Fig. \ref{massmap} and list it in Table \ref{cataloguetable}. However, it lies near an edge and a mask, so that its weak lensing characteristics are likely to be biased. We thus do not measure its mass, and will not take it into account in what follows. Cluster XLSSC025 (\cite{gavazzi}'s Cl05) is under a mask, and cannot be detected by our pipeline.

In summary, out of our 7 shapelet weak lensing detections, 4 (WL01, WL04, WL05 and WL07) have a counterpart both in our X-ray catalogue and in the KSB weak lensing catalogue by \citealt{gavazzi}, even though we remove WL07 from our subsequent analyses. One detection (WL02) appears only in our catalogue. One (WL06) has an X-ray counterpart, and appears in \cite{gavazzi}'s map, but just below the detection threshold they use for their analysis. Finally, one (WL03) has a counterpart in \citealt{gavazzi}'s catalogue, but is not selected as a C1 X-ray cluster. This proves a good agreement between the three cluster detection methods used in those observations.
X-ray clusters XLSSC005 and XLSSC029 are at too high a redshift to be detected with our surveys. Cluster XLSSC011 is too close and not massive enough to be detected by weak lensing, as will be shown in section \ref{selfct}.

\begin{figure}
\centering
\includegraphics[width=8cm,angle=0]{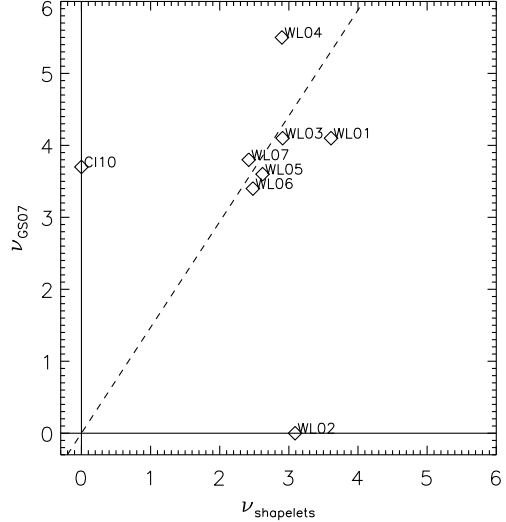} 
\caption{Comparison between our detection significances and those of \citealt{gavazzi}. Symbols are data points. Cl10 is detected by \citealt{gavazzi}, but remains invisible to our pipeline. The dashed line features the expected relation between clusters' significance in both analyses, $\nu_{\rm GS07}=1.47 ~ \nu_{\rm shapelets}$. } \label{fig_nugs}
\end{figure}

Figure \ref{fig_nugs} compares the significance of our detections with that given by \cite{gavazzi}. We make use of WL06, even though \cite{gavazzi} did not use it, but gave its significance. Given our weak lensing measurement characteristics ($n_{\rm eff}=19 ~\mbox{arcmin}^{-2}$, $\sigma_{\rm int}=0.3$) and theirs ($n_{\rm eff}=25.3 ~\mbox{arcmin}^{-2}$, $\sigma_{\rm int}=0.23$), Eq. (\ref{sf}) below allows us to compute the expected proportionality factor between our detections' significance $\nu_{\rm shapelets}$ and theirs, $\nu_{\rm GS07}$. We expect $\nu_{\rm GS07}=1.47 ~ \nu_{\rm shapelets}$. This relation is shown by the dashed line on Fig. \ref{fig_nugs}. The significance of clusters in both catalogues scale as expected. One should note that this relation depends on the measurement characteristics for both methods in those particular experiments, and should not be used as a final comparison between KSB and shapelets. More comparison on real data will be needed in order to explore this issue.

While X-ray masses listed in column (10) of Table \ref{cataloguetable}  must be taken with caution, they can be compared to the weak lensing masses listed in column (8).
Although one can notice an order of magnitude agreement between $M_{200}$(WL) and $M_{200}$(X), masses estimated from X-ray data seem slightly underestimated. This is consistent with the previously mentioned bias arising from the isothermal beta-model parametrisation.

While they do not provide strong statistics, our detections can be used to estimate $\sigma_8$, as shown below. Four detections have an X-ray counterpart and can thus be used to constrain the mass-temperature relation, provided that we add clusters from another catalogue. This is described below.

\subsection{Clusters number counts}

\subsubsection{Weak lensing selection function} \label{selfct}

A weak gravitational lensing selection function can be computed analytically (see e.g. \citealt{hamana04,marian06}) from the signal-to-noise ratio of a halo in a given cosmology and weak lensing survey parameters. We derive such a selection function, using an optimal match filter, in Berg\'e, Amara \& R\'efr\'egier (in prep).
In an observation characterised by a number density of background galaxies $n_g$, an NFW halo of convergence $\kappa$ has signal-to-noise ratio :

\begin{equation} \label{sf}
\nu = \frac{\sqrt{n_g}}{\sigma_\gamma} \sqrt{\int {\rm d}^2 x \, \kappa^2(x)}
\end{equation}
where $\sigma_\gamma$ is the r.m.s shear error per galaxy, and where we neglect projection effects and sample variance, which have been shown to have subdominant effects (\citealt{marian06}).

\begin{figure}
\centering
\includegraphics[width=8cm,angle=0]{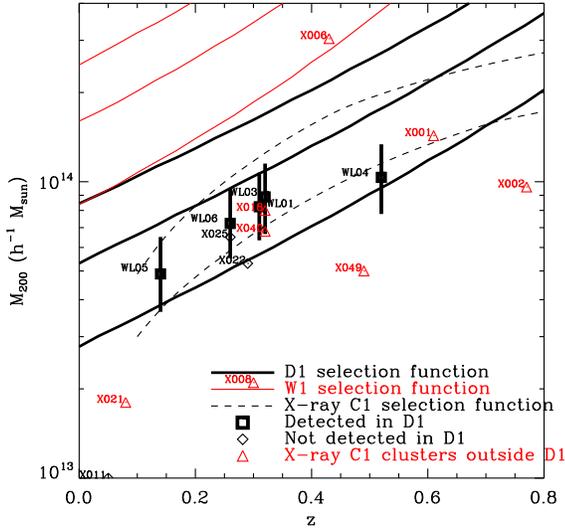} 
\caption{Weak lensing selection function for a survey like D1 (thick black ; $\sigma_{\rm int}=0.3$, $n_g$=20 arcmin$^{-1}$, redshift distribution as Eq. (\ref{eq_nz})) and W1 (red ; $\sigma_{\rm int}=0.4$, $n_g$=9 arcmin$^{-1}$, redshift distribution as Eq. (\ref{eq_nz})), assuming a WMAP3 cosmology in each case. From bottom to top, lines correspond to 2$\sigma$, 3$\sigma$ and 4$\sigma$ significance. Dashed lines show the X-ray selection function, corresponding to 50\%, and 80\% detection probability (\citealt{pacaud07} Fig. 18, lower and upper curves, respectively). Thick square symbols are our detections in the D1 data, labeled by their ID ; they are not detectable in the W1 data. Diamonds are clusters detected either by \citealt{gavazzi} or by X-ray analysis, in D1, that we do not detect for reasons listed in the text. Red triangles are C1 X-ray clusters lying outside the D1 region. 
Except for the thick square symbols (for which we use the weak lensing mass $M_{200}$(WL)), we use the X-ray mass $M_{200}$(X). Except XLSSC006 (labeled for visibility as X006), they cannot be detected by a weak lensing experiment in the W1 data. XLSSC006 is not detected because of its proximity to an edge.} \label{fig_mlim}
\end{figure}

Our selection function is shown in Figure \ref{fig_mlim} in the mass-redshift plane, for our Deep ($n_g=20$ arcmin$^{-2}$, $\sigma_\gamma=0.3$, thick black) and our Wide ($n_g=9$ arcmin$^{-2}$, $\sigma_\gamma=0.4$, red) surveys, in a cosmological model based on the three-year {\it Wilkinson Microwave Anisotropy Probe} results (WMAP3; \citealt{wmap3}), ($h,\Omega_m h^2,\Omega_b h^2,\sigma_8,w$)=(0.73,0.127,0.0223,0.76,-1). 
We use the redshift distributions for background galaxies given by equation (\ref{eq_nz}).

Figure \ref{fig_mlim} shows, from bottom to top, the minimum detectable mass for a halo at a $2\sigma$, a $3\sigma$ and a $4\sigma$ detection threshold. 
The Deep and Wide selection functions have different slopes, as a consequence of their different redshift distribution for background galaxies.
Symbols represent the position, in the redshift-mass plane, of clusters listed in Table \ref{cataloguetable}. We use the weak lensing mass $M_{200}$(WL) for WL01, WL03, WL04, WL05 and WL06 (thick square symbols) and the X-ray mass $M_{200}$(X) for other clusters (diamonds). 
Although we detect cluster XLSSC022 (WL07), we do not assess its gravitational mass, and thus show its X-ray mass in Fig. \ref{fig_mlim}.
Cluster XLSSC025 should be detectable (and is detected by \cite{gavazzi}), but it is under a mask in our analysis.
Triangle symbols (labeled $^\dagger$ in Table \ref{cataloguetable}) correspond to C1 X-ray clusters in W1 that are outside the D1 region. 
It is clear from Figure \ref{fig_mlim} that they can not be detected by our weak lensing analysis of W1. Only XLSSC006 should be seen, at the 2$\sigma$ level. However, its detection is plagued by its proximity to the edge of the image.

Also shown on Fig. \ref{fig_mlim} are the X-ray selection functions, for 50\% and 80\% detection probabilities (dashed lines) as estimated by \cite{pacaud07}.

Figure \ref{fig_mlim} shows an excellent agreement between the clusters characteristics and their predicted detectability by weak gravitational lensing.

\subsubsection{Number counts} \label{sect_counts}

\begin{figure}
\centering
\includegraphics[width=8cm,angle=0]{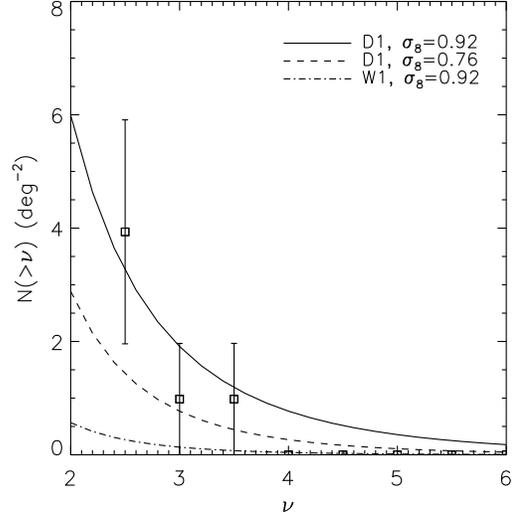} 
\caption{Cumulative cluster number density as a function of weak lensing detection significance, in the D1 data. The error bars include shot noise and sample variance. The dashed line shows the expected number counts in a WMAP3 cosmology, for the survey's characteristics. The solid line is our best fit, when varying $\sigma_8$ ($\sigma_8=0.92$). The dot-dashed line is the expected number counts for the Wide survey, with $\sigma_8=0.92$.} \label{fig_nvsnu}
\end{figure}

From equation (\ref{sf}), the expected number of haloes detected above a certain significance can be computed using a Press-Schechter approach (\citealt{press-schechter}).
For this purpose, we use the \cite{jenkins01} mass function to estimate the number of haloes that we can detect, as a function of significance threshold. Curves on figure \ref{fig_nvsnu} show such counts for different $\sigma_8$ and survey depths. \cite{miyazaki02} already used this statistic to discriminate between halo profile models.
It is used here to measure $\sigma_8$.

Most of our detections are validated by corresponding objects either in the catalogue of \citealt{gavazzi} or our X-ray C1 cluster catalogue. Nevertheless, despite its relative high significance, WL02 does not have such independent support. We indeed consider it as a false detection, and do not take it into account for cluster counts.
We then estimate the number of false detections from Monte Carlo simulations. For this purpose, we input galaxies at the position of the actual ones, but randomise their shear, and look for detections with significance higher than 2.5$\sigma$. The convergence maps that we infer from them show only false detections. We find that, in this particular experiment, we expect only one false detection above the 2.5$\sigma$ level. This is thus consistent with removing WL02. For this counting purpose, we remove WL06 and WL07 from our catalogue, since they do not reach the 2.5$\sigma$ level.
The symbols on Fig. \ref{fig_nvsnu} represent our cumulative counts, corrected from false detections. Their error bars include the effects of shot noise and sample variance, computed from Hu \& Kravtsov's (2003) analytic formula.
We then fit the expected number counts to our data as a function of $\sigma_8$, keeping all other parameters constant. In order to avoid covariance between our data points in the cumulative counts depicted by Fig. \ref{fig_nvsnu}, we performed the fit on the expected differential number counts ${\rm d}N/{\rm d}\nu (\nu)$.
We find $\sigma_8=0.92_{-0.30}^{+0.26}$ (at the 68.3\% confidence limit), for $\Omega_m=0.24$.
Despite large error bars, we can set interesting constraints thanks to the strong dependence of these counts on $\sigma_8$, as shown by the difference between the solid and dashed curves in Figure \ref{fig_nvsnu}.
The dashed line shows the expected cumulative number counts for the Deep survey in a WMAP3-like universe ($\sigma_8=0.76$). The solid line is our best fit ($\sigma_8=0.92$). The dot-dashed line shows the expected number density on the Wide survey, with $\sigma_8=0.92$.

\subsection{Mass-temperature relation}

\begin{figure}
\centering
\includegraphics[width=8cm,angle=0]{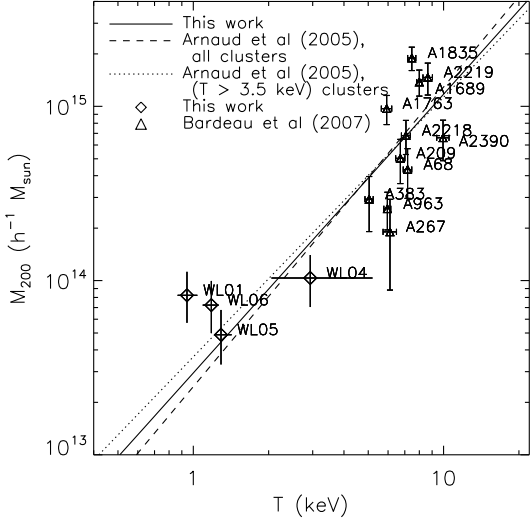} 
\caption{Mass-temperature relation, normalised to $z=0$, for our group sample (diamonds). To improve the statistics, we added clusters from \citealt{bardeau07} (triangles). We use X-ray temperature and weak lensing mass. The solid line is our best fit (Eq. \ref{eq_mt_obs}). The dashed and dotted line are \citealt{arnaud05} M-T relation, when they consider all clusters or only those with $T > 3.5 \, \mbox{keV}$, respectively.} \label{fig_mt}
\end{figure}

Under the virial equilibrium assumption, the mass and temperature of a cluster are related by the following scaling relation (\citealt{pierpaoli03})

\begin{equation} \label{eq_tstar}
\frac{M_{\rm vir}(T,z)}{10^{15} h^{-1} M_\odot} = \left( \frac{T}{T_*} \right)^{3/2} \left[\Delta_c(z) E(z)^2\right]^{-1/2} \left[1-2 \frac{\Omega_\Lambda(z)}{\Delta_c(z)} \right]^{-3/2}
\end{equation}
where $M_{\rm vir}$ is the virial mass, $T$ is the virial temperature, $T_*$ is a normalisation factor, and $E(z)^2=\Omega_m(1+z)^3+\Omega_\Lambda+\Omega_k(1+z)^2$.
$\Delta_c(z)$ is the overdensity inside the virial radius, in units of the critical density. We compute it using the fitting formula by \cite{wk03} for $\Delta_{\rm vir} = \Delta_c/\Omega_m$, which is very similar to an earlier approximation by \cite{nakamura96} given by \cite{ks96} for a universe with arbitrary $\Omega_m$.

A more general relation often used to fit observations makes use of a related normalisation factor $M_*$ and is given, at redshift $z=0$, by 

\begin{equation} \label{eq_mt}
M_{200} \approx  M_* \left( \frac{T}{4 \,\, {\rm keV}} \right)^\alpha
\end{equation}
where $M_{200}$ is the mass inside the sphere of mean overdensity 200 times higher than the critical density and $\alpha=3/2$ in the hydrostatic equilibrium assumption (e.g. \cite{arnaud05}). 
Hereafter, to account for redshift evolution, we normalise all our temperatures to $z=0$ by dividing them by $E(z)^{2/3}$.

Measuring $\sigma_8$ from X-ray counts is affected by the degeneracy $\Omega_m^{0.6} \sigma_8 \propto T_*^{-0.8}$ (\citealt{pierpaoli03}). \cite{pierpaoli03} have shown that the uncertainty in $M_*$ is the main concern in measurements of $\sigma_8$ from X-ray cluster observations alone. Such data is limited by the requirement that the cluster masses be inferred from the X-ray profiles.
\cite{smith03} have also shown that unrelaxed clusters, being hotter than relaxed clusters, provide a supplementary bias to the $\sigma_8$ estimate.
It is thus important to have a mass estimate independent of the hydrostatic equilibrium assumption. 
Weak gravitational lensing gives such an estimate. Combined with X-ray temperature, it can be efficiently used to constraint the M-T relation, independently of the cluster physical state. \cite{hjorth98,pedersen06,bardeau07} have already used it to measure the M-T relation normalisation. 

As described above, we have the weak lensing mass and X-ray temperature of only four groups.
Hence, to increase our statistics, we add \cite{bardeau07}'s clusters to our catalogue, providing us with 11 additional clusters. \cite{bardeau07} estimated cluster masses by fitting an NFW model to their tangential shears. We should note here that since \cite{bardeau07}'s mass estimation and ours are based on different techniques, our subsequent analysis of the mass - temperature relation could be slightly biased due to possible calibration differences.
Figure \ref{fig_mt} shows the relation between the temperature and the weak lensing mass $M_{200}$ for the combined catalogues. Diamonds are our groups, labelled WL{\it id}, triangles are \cite{bardeau07}'s clusters, labelled A{\it id}. 
\cite{bardeau07} proceeded to the weak lensing analyses of massive haloes, the temperature of which were obtained by \cite{zhang07} and \cite{ota04} ; particularly, they estimated their weak lensing mass and measured the scale relations for those clusters.
They fitted their sample by varying both $\alpha$ and $M_*$, and found a large slope, far from the hydrostatic equilibrium assumption, $\alpha=4.6 \pm 0.7$. 
Doing the same analysis on the larger range in mass that the addition of both catalogues probes, from galaxy groups to galaxy clusters, we find :

\begin{equation} \label{eq_mt_obs}
\frac{M_{200}}{10^{14} h^{-1} M_\odot} = 2.71_{-0.61}^{+0.79} \left( \frac{T}{4 \,\, \mbox{keV}} \right) ^{1.60 \pm 0.44},
\end{equation}
which is in good agreement with \cite{arnaud05} (whether they use all clusters or only the most massive ones), \cite{bardeau07} (when they fix $\alpha=1.5$), \cite{pedersen06} or \cite{hoekstra07}. 
The solid line on Fig. \ref{fig_mt} is our best fit. The dashed line is the best fit from \cite{arnaud05}, when they consider all clusters in their catalogue. The dotted line is their best fit when they consider their most massive clusters ($T > 3.5 \,\, \mbox{keV}$).

Our result is consistent with self-similarity evolution for galaxy clusters down to low temperatures. It is also consistent with previous measurements which observe a steepening of the M-T relation at the low mass end, due to the expected self-similarity breaking for such masses (e.g. Nevalainen et al. 2000; Finoguenov et al. 2001;  \citealt{arnaud05}). Moreover, one must be aware that the galaxy groups we consider were detected just above our weak lensing selection function (Fig. \ref{fig_mlim}). Due to the expected scatter in the M-T relation, those groups can represent only the most massive ones with temperature ranging about 1 keV. Our group sample could thus bias our fit towards a flat slope for the M-T relation. The analysis of more low temperature groups will be needed to further explore this issue.


\begin{figure}
\centering
\includegraphics[width=8cm,angle=0]{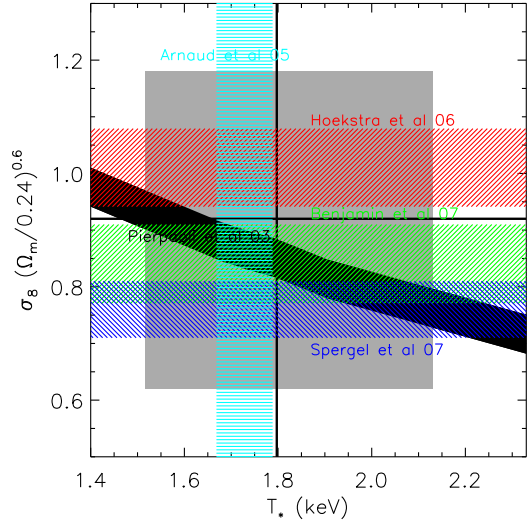} 
\caption{Domain allowed for by different measurements, in the $T_*$ - $\sigma_8 (\Omega_m / 0.24)^{0.6}$ plane. 
The shaded region shows the constraints given by our $\sigma_8$ and $T_*$ measurements.
Thick lines are our best fits. The slanted black region correspond to the $1 \sigma$ constraints on the $\Omega_m^{0.6} \sigma_8 \propto T_*^{-0.8}$ relation from \citealt{pierpaoli03}. The vertical, light blue, shaded region shows the $1\sigma$ error on $T_*$ from \citealt{arnaud05}. CMB derived constraints of $\sigma_8$ (\citealt{wmap3}) are shown by the horizontal dark blue shaded region. Cosmic shear $\sigma_8$ estimations from \citealt{hoekstra06} and \citealt{benjamin07} are marked by the red and green horizontal shaded regions. \citealt{hoekstra06}'s constraints are typical of cosmic shear results. They are higher than X-ray estimations, marked by the intersection between \citealt{pierpaoli03}'s and \citealt{arnaud05}'s allowed domains.} \label{fig_constraints}
\end{figure}

\section{Discussion} \label{discussion}

The power spectrum normalisation $\sigma_8$ has been measured with different probes, such as X-ray clusters of galaxies, CMB, and cosmic shear (i.e. statistics of weak gravitational lensing). Some discrepancies have emerged between the preferred value from those measurements. Recent CMB observations favour a low $\sigma_8$ and cosmic shear used to emphasise a high value (see e.g. \citealt{alexreview} for a review). X-ray clusters provide intermediate measurements (see e.g. \citealt{pierpaoli03} for a review). The dominant discrepancy between cosmic shear and X-ray clusters has recently been reduced by \cite{jarvis06}, who measured $\sigma_8 \approx 0.81$ for $\Omega_m=0.26$ when using cosmic shear alone, followed by \cite{benjamin07}, who used \cite{ilbert06}'s improved galaxies photometric redshifts,  and measured $\sigma_8=0.84$ for $\Omega_m=0.24$, and \cite{fu08} who found a consistent value.
To clarify these discrepancies, one needs to measure both the power spectrum and the M-T relation normalisations $\sigma_8$ and $M_*$, as we discuss here.

The regions allowed for by different measurements on the $T_*$ - $\sigma_8 (\Omega_m / 0.24)^{0.6}$ plane are shown by Figure \ref{fig_constraints}. The shaded region shows the constraints given by our $\sigma_8$ and $T_*$ measurements. Our best fits are shown by the thick lines.
The slanted black band on the figure is the 68.3\% bound on the $\Omega_m^{0.6} \sigma_8 \propto T_*^{-0.8}$ relation from \cite{pierpaoli03} using X-ray clusters.
Its intersection with the vertical light blue band (\citealt{arnaud05}'s $T_*$ estimation) gives the current value for $\sigma_8$ favoured by X-ray cluster observations, $\sigma_8 \approx 0.77 \pm 0.06$ for an $\Omega_m=0.3$ universe (e.g. \citealt{pierpaoli03}), which corresponds to $\sigma_8 (\Omega_m / 0.24)^{0.6} \approx 0.88 \pm 0.05$. 
This value is higher than that measured by \cite{wmap3} from CMB analyses of WMAP3 (dark blue), but lower than most cosmic shear analysis, like that of \cite{hoekstra06} made with CFHTLS Wide data (red). This highlights the discrepancy between X-ray and weak lensing estimates of $\sigma_8$ mentioned above. However, \cite{benjamin07} give a lower estimate for $\sigma_8$, consistent with X-ray measurements (green). This could be a sign that other cosmic shear analyses did not take some systematics into account, and have thus overestimated $\sigma_8$. 
According to \cite{benjamin07}, previously published analyses made use of insufficiently-accurate galaxy photometric redshifts. Using \cite{ilbert06}'s redshifts yielded a lower value of $\sigma_8$ both for cosmic shear (\citealt{benjamin07}) and for our cluster count analysis. We found a 5\% decrease in our $\sigma_8$ estimation when going from previous redshift distributions to \cite{ilbert06}'s ones. This is less than the change reported by \cite{benjamin07}, and our best fit still tends to favour a higher value for $\sigma_8$, but is limited by low statistics.
\cite{smith03} have analysed the bias from unrelaxed clusters in $\sigma_8$ measurement using lensing clusters and the M-T relation. They found that unrelaxed clusters are 30\% hotter than relaxed clusters : using unrelaxed clusters can provide 20\% overestimates of $\sigma_8$. This is enough to explain the large range of measured $\sigma_8$, from $\approx$ 0.6 to $\approx$ 1. They estimated $\sigma_8 (\Omega_m / 0.24)^{0.6} = 0.86 \pm 0.23$.
Estimates from X-ray alone can also be affected by systematics, such as the mass estimate from X-ray profiles of clusters. 
For example, a slight decrease of $T_*$ would cause an increase of the X-ray estimate for $\sigma_8$. A better insight into this will come from an accurate measurement of $T_*$, preferably with mass estimation methods independent of cluster physics. 
Large combined weak lensing and X-ray surveys will be needed to disentangle the situation. They will provide both independent constraints on $\sigma_8$, and insights on $T_*$.

\section{Prospects for future surveys} \label{impact}

In the following, we investigate the impact of future combined blind weak lensing and X-ray surveys on the measurement precision of the power spectrum and the mass-temperature relation normalisations. We take the WMAP3 (\citealt{wmap3}) cosmology as our fiducial model. We consider two different ground based survey strategies for our weak gravitational lensing analysis : deep and wide surveys similar to the CFHTLS Deep and Wide surveys. We use their observed weighted number density of useful background galaxies to be $n_g=20$ arcmin$^{-2}$ and $9$ arcmin$^{-2}$, respectively, distributed according to Eq. (\ref{eq_nz}). We also assume the intrinsic ellipticity and shape measurement error to be $\sigma_\gamma=0.3$ in both cases. 
Following the CFHTLS scheduling, we take for exposure times 40 hours per square degree for the deep survey and 1 hour per square degree for the wide survey.

\subsection{$\sigma_8$ measurements}

\begin{figure}
\centering
\includegraphics[width=8cm,angle=0]{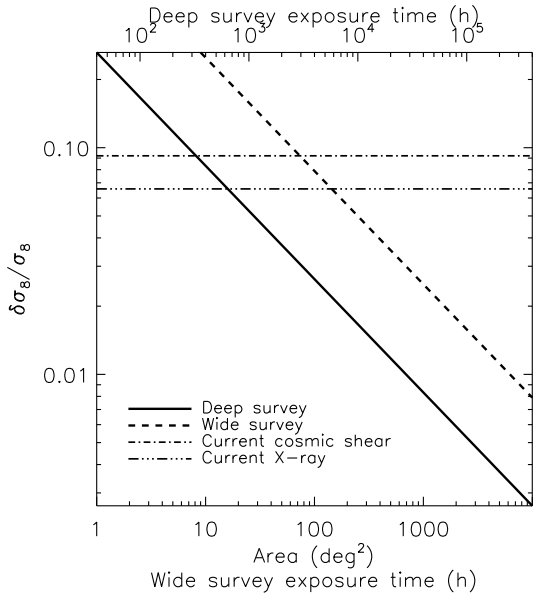} 
\caption{Relative errors on $\sigma_8$, from clusters counts in a weak gravitational lensing survey, as a function of survey size and integration time. All other parameters are kept constants. We assume that 1 deg$^2$ of wide requires 1 hour of observation time, and 1 deg$^2$ of deep requires 40 hours of observation time.
That is, the lower x-axis shows area as well as the wide survey exposure time ; the upper x-axis shows the deep survey exposure time. The thick solid line corresponds to a deep survey, and the thick dashed line to a wide survey. The flat lines show the current error measurement from cosmic shear statistics (dash-dot, Hoekstra et al. 2006; Benjamin et al. 2007), and from X-ray clusters (dash-dot-dot, Pierpaoli et al. 2003; APP05).} \label{fig_prosp1}
\end{figure}

We first investigate the impact of future surveys on the $\sigma_8$ measurement.
Using the Press-Schechter approach described in section \ref{sect_counts}, we estimate the number of weak lensing detections with significance higher than 2.5, taking into account shot noise and sample variance. 
We assume that all clusters have a spherically symmetrical NFW profile. We thus neglect the effect of haloes' asphericity shown by Clowe et al. (2004) : triaxial haloes oriented along the line of sight appear more massive than triaxial haloes of the same mass, but perpendicular to the line of sight, and thus have a higher signal-to-noise ratio. Clowe et al. (2004) have shown that this approximation does not yield any difference in the mass measurement dispersion. 
Figure \ref{fig_prosp1} shows the 68.3\% relative error on $\sigma_8$ that can be reached by counting weak lensing detected clusters as a function of their significance, for a deep (thick solid line) and a wide (thick dashed line) surveys, as a function of survey's size and observing time. Because of the higher number density of clusters it allows one to detect, a deep survey provides errors 3 times lower than a wide survey of the same size. However, for a given exposure time, a wide survey provides errors 2.1 times lower than a deep one. That means that the gain due to the coverage (and detectable clusters number) increase is faster than the one due to depth increase. A larger coverage is also advantageous in that it makes sample variance fall down rapidly. 
Moreover, a wide survey detects the most massive haloes, the physics of which is better understood. 
Consequently, in a survey strategy driven by exposure time, one should prefer a wide survey. The flat dashed-dot line shows the current constraints provided by cosmic shear analyses (\citealt{hoekstra06,benjamin07}). The flat dashed-dot-dot line shows the current constraints from the combination of X-ray $M_*$ measurement (\citealt{arnaud05}) and X-ray cluster counts (\citealt{pierpaoli03}). 

Detecting and counting clusters on a 10 deg$^2$ deep survey will be competitive with current cosmic shear measurements, whereas 20 deg$^2$ of coverage is needed to compete with current X-ray clusters measurements. Those figures transform as 100 deg$^2$ and 200 deg$^2$ for a wide survey. That is, to compete with current cosmic shear surveys, one needs 400 hours of deep survey exposure, or 100 hours of wide survey exposure. Double these times are required to compete with X-ray surveys. A wide survey, less demanding in exposure time than a deep one, should then be used. Counting clusters on the entire planned CFHTLS Wide Survey 170 deg$^2$ will provide a 6\%  fractional error on the $\sigma_8$ measurement. Reaching the 1\% fractional error will require a 7000 deg$^2$ wide survey, or a 700 deg$^2$ deep survey. Future surveys (e.g. Pan-STARRS\footnote{http://panstarrs.ifa.hawaii.edu}, LSST\footnote{http://www.lsst.org}, DUNE\footnote{http://www.dune-mission.net}) will be able to achieve such errors.

\begin{figure}
\centering
\includegraphics[width=8cm,angle=0]{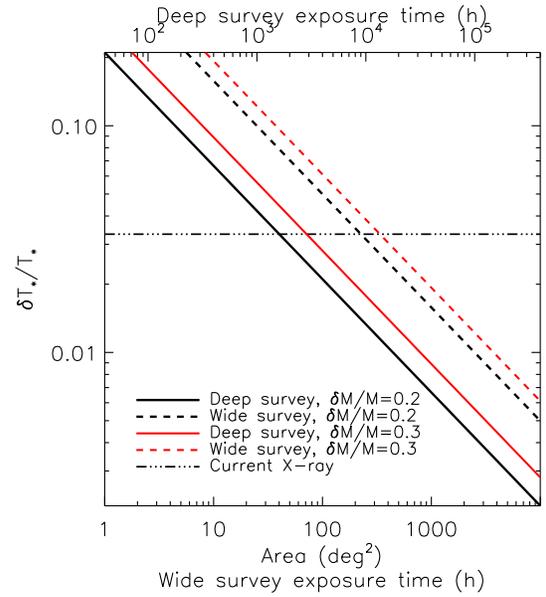} 
\caption{Relative errors on $T_*$ from combined weak gravitational lensing and X-ray surveys, as a function of survey size and integration time. All other parameters are kept constants. The lower x-axis shows area as well as the wide survey exposure time; the upper x-axis shows the deep survey exposure time. Solid lines show the errors for a deep survey, assuming the fractional error on weak lensing mass measurement is 20\% (thick black) and 30\% (red). Dashed lines, with the same colour indexing, show the errors for a wide survey. The current error measurement from X-ray clusters is shown by the flat dashed-dot-dot line (Pierpaoli et al. 2003; APP05). We made the same assumption about the relation between survey area and observation time as in Fig. \ref{fig_prosp1}.} \label{fig_prosp2}
\end{figure}

\subsection{$T_*$ measurements}

We now turn to the precision that can be reached on $T_*$ measurements by future joint surveys. What matters here is not cluster counts as a function of significance, but as a function of mass.
As we see from figure \ref{fig_mlim}, low mass clusters of galaxies cannot be seen through weak gravitational lensing since they do not create high enough signal-to-noise ratios. A deep survey captures lower mass clusters than a wide survey, but deep and wide surveys give access to the same number of massive clusters.
Therefore, a deep survey is naturally focused on the physics of galaxy groups (e.g. it can probe similarity breaking at the low mass end of the mass-temperature relation). A wide survey gives the same statistics on massive haloes, generally used to measure the mass-temperature relation normalisation : for this purpose, one should then choose a wide survey. 
To compare the merits of both deep and wide survey on the $T_*$ estimation's precision, we simulate M-T relations for both types of survey. We take a realistic scatter into account, $\sigma_{\rm log,int}=0.051$ for the logarithmic M-T relation (\citealt{arnaud05}). We assume that masses are measured through weak gravitational lensing. We measure $T_*$ and the error on its estimate, by assuming (1) that our cluster sample is complete, (2) that we only make use of those clusters detected in our blind survey, and (3) that we know the X-ray temperature of each of them.
We also investigate the influence of the mass estimation fractional error $\delta M/M$.

Figure \ref{fig_prosp2} shows the 68.3\% error on $T_*$ that can be reached from a combined blind X-ray and deep (solid line) or wide (dashed line) weak gravitational lensing surveys, as a function of survey's size and integration time. Here again, as for the error on the power spectrum normalisation, a deep survey gives errors 2.3 times lower than a wide one with the same sky area coverage. On the other hand, a wide survey gives errors 2.7 times lower than a deep one with the same exposure time. 
The dependence on area underlines the $T_*$ estimation's reliance on the number of useable haloes for the M-T relation fitting.
Figure \ref{fig_prosp2} also shows the sensitivity of the $T_*$ estimation to the mass measurement errors. The black lines assume $\delta M / M =0.2$, and the red ones $\delta M/M=0.3$, which are the current fractional errors on mass measurement from weak lensing. Going from $\delta M/M=0.3$ to $\delta M/M=0.2$ allows one to reduce the error on $T_*$ by a factor of 1.3 (resp. 1.2) for a deep (resp. wide) survey of a given sky area.
The flat solid line represents the current error on $T_*$ from X-ray clusters (\citealt{arnaud05}). Assuming a 20\% error measurement on weak lensing masses, one needs a 50 (resp. 300) deg$^2$ weak lensing deep (resp. wide) survey to reach the current error. 
Reaching the 1\% fractional error (for our fiducial model with $T_*=1.9$) will require a 2500 deg$^2$ wide survey, or a 500 deg$^2$ deep survey. Weak lensing surveys like LSST or DUNE combined with X-ray surveys like eROSITA will be able to reach such a limit.

In a survey strategy driven by exposure time, a wide survey of 2500 deg$^2$ (2500 hours) will be able to reach the 1\% accuracy both on $\sigma_8$ and $T_*$, at a much cheaper expense than a deep survey. Nevertheless, a deep survey will still be useful to probe high redshift regions ($z \geqslant 0.8$), and to study low mass clusters of galaxies ($M \leqslant 10^{14} h^{-1} M_\odot$).


\section{Conclusion} \label{conclusion}

We have presented the first shapelet analysis of weak gravitational lensing surveys. We have constructed convergence maps of the CFHTLS Deep D1 field, and of 4 deg$^2$ of the CFTHLS Wide W1 field, which include the D1 field. We have detected six clusters of galaxies, through the lensing signal they generate. Our D1 map is in good agreement with that of \cite{gavazzi}, precedently created using the KSB shear measurement method. 
We combined our weak lensing data with the X-ray analysis of XMM-LSS C1 clusters lying in the same region of the sky (\citealt{pacaud07}).
These three clusters catalogue are consistent. All our shapelet detections have either an X-ray counterpart or a KSB detection. 
Counting our detections and accounting for the weak lensing selection function allowed us to constrain the power spectrum normalisation $\sigma_8 (\Omega_m / 0.24)^{0.6} = 0.92_{-0.30}^{+0.26}$. 
The combination of lensing masses and X-ray temperatures provided us with a new measurement of the mass-temperature relation normalisation $T_*$ (or equivalently $M_*$) for clusters of galaxies, $M_* = 2.71_{-0.61}^{+0.79} \, 10^{14} h^{-1} M_\odot$. Our results, though limited by low statistics and sample variance, are consistent with other current estimates. We also measured the slope of the mass-temperature relation, and found it consistent with self-similarity for low mass clusters, $\alpha=1.60 \pm 0.44$.
We have shown that one must measure both $\sigma_8$ and $T_*$ from combined weak lensing and X-ray surveys to investigate the discrepancy between independent measurements of $\sigma_8$ from different probes.

Weak lensing surveys are becoming more and more effective, and are currently being optimised for best extracting cosmological information.
Optimal surveys will allow us to provide more accurate estimates of $\sigma_8$ and $T_*$, and to disentangle the current $\sigma_8$ issue (\citealt{amara06}).
We have compared the merits of weak lensing deep and wide blind surveys, based on the CFHTLS, at estimating $\sigma_8$. We also looked at their merits at estimating $T_*$ while combined with an X-ray survey on their region of the sky. We found that for experiments driven by exposure time constraints, a wide survey will give $\approx$ 3 times lower errors on the estimates of both $\sigma_8$ and $T_*$. 
To secure the measurement of $\sigma_8$ and $M_*$ with the current statistical accuracy, a 200 deg$^2$ and a 300 deg$^2$ wide surveys will be needed respectively.
We finally found that a 7000 deg$^2$ wide survey will be able to reach the 1\% accuracy both on the power spectrum and mass-temperature relation normalisations.

\section*{Acknowledgements}
The authors wish to thank Herv\'e Aussel, Krys Libbrecht, Jean-Baptiste Melin, Yannick Mellier, Sandrine Pires, Trevor Ponman, Jean-Luc Starck, Genevi\`eve Soucail and Romain Teyssier for useful discussions, and Cathy Horellou for comments on the first version of the paper.


\label{lastpage}

\end{document}